\documentclass[a4paper,english,10pt,twoside]{article}
\usepackage[english]{babel}
\usepackage[latin1]{inputenc}
\usepackage{amsmath}
\usepackage{graphicx}

\usepackage{fancyhdr}
\fancypagestyle{plain}{%
  \fancyhf{}%
  \fancyhead[C]{\small\it{Twenty years of giant exoplanets - Proceedings of the Haute Provence Observatory Colloquium, 5-9 October 2015\\ Edited by I. Boisse, O. Demangeon, F. Bouchy \& L. Arnold} }
  \fancyfoot[C]{\normalsize \thepage}%
}

\newcommand{\nc}   {\newcommand}
\nc{\dsfrac}[2]{{\displaystyle\frac{#1}{#2}}}

\newcommand{\Jcal}  {{\EuScript J}}

\nc{\ergs}  {\mathrm{erg}/\mathrm{gm\;s}}
\nc{\ergc}  {{\rm erg}~{\rm cm}^{-3}}
\nc{\ergcs} {{\rm erg}~{\rm cm}^{-2}~{\rm s}^{-1}}
\nc{\ms}   {{\rm m}~{\rm s}^{-1}}

\nc{\Eem}   {\widetilde{E}_\mathrm{e}}
\nc{\Eexm}  {\widetilde{E}_\mathrm{ex}}
\nc{\Eim}   {\widetilde{E}_\mathrm{I}}
\nc{\ER}    {E_\mathrm{R}}
\nc{\FL}    {F_\mathrm{Ly}}
\nc{\FR}    {F_\mathrm{R}}
\nc{\FRa}   {F_\mathrm{R,1}}
\nc{\FFR}   {\mathbf{F}_\mathrm{R}}
\nc{\Fe}    {F_\mathrm{e}}
\nc{\FFe}   {\mathbf{F}_\mathrm{e}}
\nc{\Fsat}  {q_\mathrm{sat}}
\nc{\Ke}    {\EuScript{K}_\mathrm{e}}
\nc{\mfp}   {l_\mathrm{e}}
\nc{\mrc}   {\mathrm{c}}
\nc{\mrr}   {\mathrm{r}}
\nc{\nH}    {n_\mathrm{H}}
\nc{\nel}   {n_\mathrm{e}}
\nc{\Pe}    {P_\mathrm{e}}
\nc{\Pg}    {P_\mathrm{g}}
\nc{\PR}    {P_\mathrm{R}}
\nc{\Qelc}  {Q_\mathrm{elc}}
\nc{\Qinc}  {Q_\mathrm{inc}}
\nc{\gyr}   {r_\mathrm{B}}
\nc{\Ta}    {T_a}

\nc{\divFe} {\nabla\cdot\FFe}
\nc{\divFR} {\nabla\cdot\FFR}
\nc{\FLj}   {F_\mathrm{Ly,\Jcal}}
\nc{\FRj}   {F_\mathrm{R,\Jcal}}

\nc{\aap}   {A\&A}
\nc{\aaps}   {A\&AS}
\nc{\apj}   {ApJ}
\nc{\mnras}   {MNRAS}
\nc{\aapr} {A\&ARv}
\nc{\pasp} {PASP}
\nc{\apjl} {ApJL}
\usepackage{txfonts}
\usepackage[squaren,cdot]{SIunits}
\usepackage{tabularx}
\usepackage[breaklinks=true]{hyperref} 
\usepackage{natbib,twoopt}
\bibpunct{(}{)}{;}{a}{}{,} 
\makeatletter
\newcommandtwoopt{\citeads}[3][][]{\href{http://adsabs.harvard.edu/abs/#3}%
{\def\hyper@linkstart##1##2{}%
\let\hyper@linkend\@empty\citealp[#1][#2]{#3}}}
\newcommandtwoopt{\citepads}[3][][]{\href{http://adsabs.harvard.edu/abs/#3}%
{\def\hyper@linkstart##1##2{}%
\let\hyper@linkend\@empty\citep[#1][#2]{#3}}}
\newcommandtwoopt{\citetads}[3][][]{\href{http://adsabs.harvard.edu/abs/#3}%
{\def\hyper@linkstart##1##2{}%
\let\hyper@linkend\@empty\citet[#1][#2]{#3}}}
\newcommandtwoopt{\citeyearads}[3][][]%
{\href{http://adsabs.harvard.edu/abs/#3}
{\def\hyper@linkstart##1##2{}%
\let\hyper@linkend\@empty\citeyear[#1][#2]{#3}}}
\makeatother

\makeatletter 

\def\@maketitle{%
  \vskip 2em%
  \begin{center}%
  \let \footnote \thanks
    {\LARGE\textbf \@title \par}%
    \vskip 1.5em%
    {\normalsize
      \lineskip .5em%
      \begin{tabular}[t]{c}%
        \@author
      \end{tabular}\par}%
    \vskip 1em%
    {\normalsize \@date}%
  \end{center}%
  \par
  \vskip 1.5em}
\makeatother

\newcommand{\affil}[1]{\small{\hskip-0.55cm #1}}
\newcommand{\acknowledgments}[1]{\small{ \vskip3mm \hskip-0.55cm {Acknowledgments: #1}}}

\begin{document}

\setcounter{page}{1}  

\title{Hot Jupiters and Super-Earths} 
\author{ A.\ J.\ Mustill$^{1}$, M.\ B.\ Davies$^1$, A.\ Johansen$^1$} 
\date{} 
\maketitle
\affil{  $^1$Lund Observatory, Department of Astronomy \& Theoretical Physics, Lund University, Box 43, SE-221 00 Lund, Sweden (\texttt{alex@astro.lu.se})
}

\vskip1cm

\begin{abstract}
We explore the role of dynamics in shaping planetary system multiplicities, focussing on two 
particular problems. \textbf{(1)} We propose that the lack of close-in super-Earths in hot 
Jupiter systems is a signature of the migration history of the hot Jupiters and helps to 
discriminate between different mechanisms of migration. We present N-body simulations of 
dynamical migration scenarios where proto-hot Jupiters are excited to high eccentricities 
prior to tidal circularisation and orbital decay. We show that in this scenario, the 
eccentric giant planet typically destroys planets in the inner system, in agreement with the 
observed lack of close super-Earth companions to hot Jupiters. \textbf{(2)} We explore the role 
of the dynamics of outer systems in affecting the multiplicities of close-in systems such as 
those discovered by \textit{Kepler}. We consider specifically the effects of planet--planet 
scattering and Kozai perturbations on an exterior giant planet on the architecture of the inner 
system, and evaluate the ability of such scenarios to reduce the inner system's 
multiplicity and contribute to the observed excess of single \textit{Kepler} planets.
\end{abstract}

\section{High-eccentricity migration of hot Jupiters destroys inner planets}

The formation and migration of hot Jupiters has been discussed since their discovery. 
Migration may take place at early times on a near-circular orbit in a protoplanetary disc 
\citep{Lin+96,Ward97}. Alternatively, migration may occur after protoplanetary disc 
dissipation, as the giant planet's orbital eccentricity is excited to high values and 
then the orbit is re-circularised by tidal forces, shrinking the planet's semi-major axis 
in the process \citep{RF96}. Several means of eccentricity excitation are possible, 
such as planet--planet scattering \citep{RF96,Chatterjee+08} and various classes of 
secular perturbation 
\citep{WuMurray03,FabryckyTremaine07,WuLithwick11,BeaugeNesvorny12,Petrovich15a,Petrovich15b}.

The dynamical architecture of hot Jupiter systems may help to distinguish these two 
migration pathways. The high-eccentricity migration route requires that the hot Jupiters have, 
or previously had, massive planetary or stellar companions on wide orbits, which appears to 
be supported by recent observations, with around 70\% of hot Jupiters having such companions 
\citep{Knutson+14,Ngo+15}. In contrast, the frequency of low-mass, close-in companions to hot 
Jupiters appears to be very low. No such companions have been discovered by RV surveys, while 
none of the hot Jupiters in the original \textit{Kepler} field has low-mass transiting planets, 
nor do they show TTVs that these planets may induce \citep{Steffen+12}. The sole exception to 
this pattern is WASP-47, a system where the giant planet \citep{Hellier+12} pairs up with an additional 
wide-orbit giant \citep{Neveu+15} and also possesses two low-mass companions discovered by K2 transit 
photometry \citep{Becker+15}.

\begin{figure}[t]
  \centering
  \hspace{-2cm}
  \includegraphics[width=12cm]{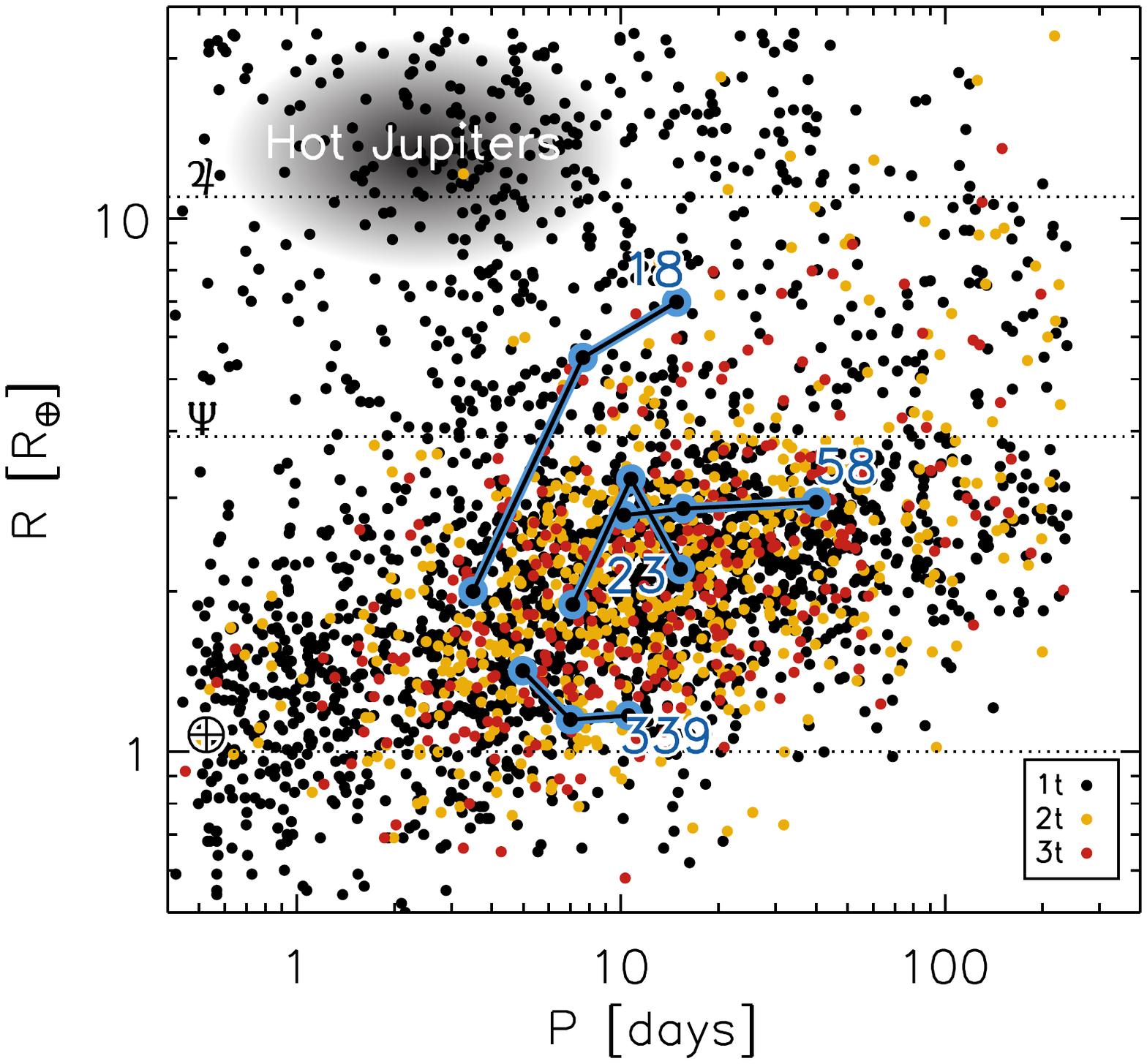}
  \caption{Periods and radii of \textit{Kepler} single-, double- and triple-candidate systems. Hot 
  Jupiters are single (top left of plot). Inner systems used as templates for further study are 
  marked in blue.}
  \label{fig:kepler}
\end{figure}

At present then, hot Jupiters appear to only rarely possess close companions (Figure~\ref{fig:kepler}). 
This is despite the fact 
that small, close-in planets are very commonly observed both by \textit{Kepler} transit photometry 
\citep{Fressin+13} and RV surveys \citep{Howard+10}. If the formation and migration of hot Jupiters 
and the formation of close-in planets are uncorrelated processes, one would therefore expect that 
around half of all migrating hot Jupiters would interact with close-in planets \textit{en route} to 
their final destination. Previous work has shown that migration through the gas disc is not completely 
efficient at destroying other planets or suppressing their formation 
\citep[e.g.,][]{Mandell+07,FoggNelson09,Ketchum+11,Ogihara+14}. Here we concentrate on the case of 
high-eccentricity migration, showing that during its phase of high eccentricity a giant that will become 
a hot Jupiter is almost guaranteed to destabilise and destroy any small planets close to the star. Full 
details of our work can be found in \cite{Mustill+15}.

\begin{figure}[t!]
  \centering
  \includegraphics[width=\textwidth]{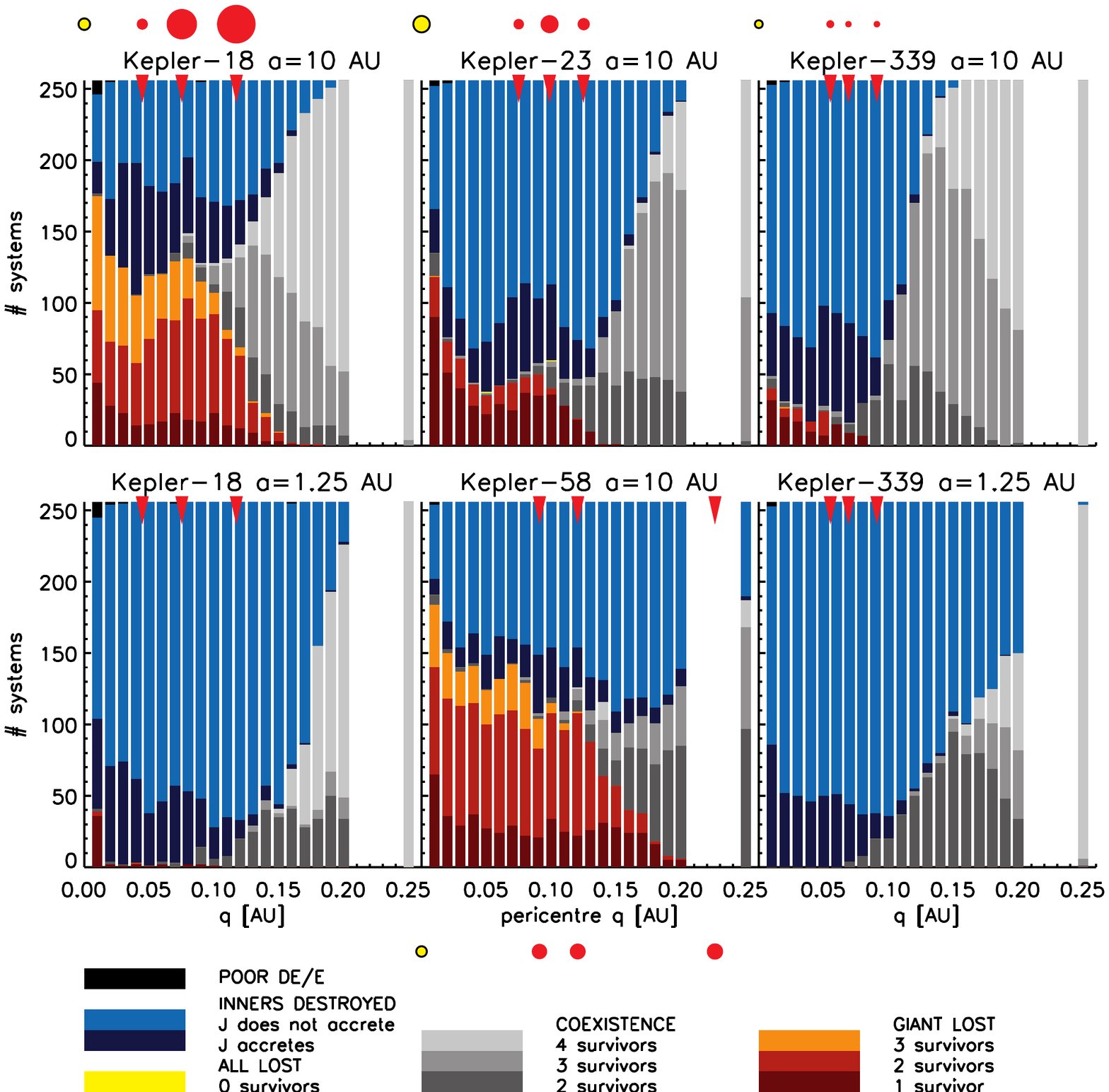}
  \caption{Stacked bar chart showing the effects
    of a highly-eccentric proto-hot Jupiter encountering inner planetary systems, for a
    variety of inner systems (shown above and below the bars),
    giant planet pericentres and semi-major axes. Each bar shows the outcome of
    a set of 256 simulations. Blue shows systems where the inner planets were destroyed, red/orange where the giant
    was lost and at least one inner planet survived. Grey shows where the giant and at least one inner planet
    survive for the 1Myr integration. Where the giant's pericentre is sufficiently small to permit later tidal
    circularisation ($\lesssim0.04$ au), coexistence is highly unlikely and either the inner planets are destroyed
    (usually forced into the star) or the giant itself is lost (usually to ejection).}
  \label{fig:bars}
\end{figure}

\subsection{Results}

We conducted a series of N-body integrations using the BS algorithm in the \textsc{Mercury} package 
\citep{Chambers99}. Our set-up was to take a \textit{Kepler} triple-planet system on circular orbits 
(Kepler-18, Kepler-23, 
Kepler-58, Kepler-339; these roughly span the range of planetary radii of \textit{Kepler} planets) and 
add to it an eccentric giant planet on a wide orbit. We explored pericentres of the giant from 0.01 au 
to 0.25 au, and semi-major axes from 1.25 au to 10 au. Hence, we ignore the evolution prior to 
eccentricity excitation---be it planet--planet scattering, Kozai perturbations from a binary or secular 
planet--planet interactions---and focus on the effects on the inner planets if that eccentricity is 
imposed at the beginning of the integration. The giant planets' inclinations are assigned isotropically 
since mechanisms of exciting eccentricity often excite high inclinations too, while the inclinations of 
the inner planets are within $5^\circ$ of the reference plane. Giant planets are released from apocentre.

The results are summarised in Figure~\ref{fig:bars}. There are three broad classes of outcome: (1) The giant planet 
destroys the system of inner planets, usually forcing them into the star. More rarely the giant collides with 
one or more of the inner planets, which may lead to an observable enrichment of the planet's core. (2) The giant 
planet is ejected by the inner planets. In this case the number of inner planets may be reduced as they are destabilised 
and merge from three to two or one survivor. Ejection of the giant is possible because when coming in from a very 
large apocentre, only a small kick at pericentre is needed to increase the apocentre still further, while the more 
tightly-bound inner planets do not experience such strong effects. Ejection is more common when the giant has a wider 
orbit or when the inner planets are comparatively more massive. (3) The giant and at least one inner planet coexist 
after the 1Myr integration ends. This typically happens when the giant's orbit is not initially overlapping any of the 
inner planets', and almost never happens when the giant's pericentre is small enough to permit subsequent tidal 
circularisation ($q\lesssim0.04$ au).

The pericentres of the giants vary very little during their interaction with the inner planets, while their 
semimajor axes may change significantly. Many of our surviving giants will form hot Jupiters given sufficient 
time for tidal circularisation of their orbits, since their pericentres remain small. Giants with larger pericentres 
($\gtrsim0.1$ au) may still experience significant changes to their semimajor axes as they destroy the inner 
planets, with their orbits sometimes shrinking to $\sim0.5$ au. In doing so they maintain small pericentres and high 
eccentricities (up to $\sim0.8$). This is a region hard to populate through \textit{in situ} scattering of giant 
planets \citep{Petrovich+14}, and these systems may have in the past had close-in super-Earths or Neptunes.

Our simulations show that the formation of systems similar to WASP-47, 
where the hot Jupiter ($P=4.16$ d) is accompanied by low-mass planets on interior 
($P=0.79$ d) and exterior ($P=9.03$ d) orbits, is almost impossible under high-eccentricity migration. 
At the end of our $\sim30\,000$ integrations we find that 23 of the surviving giants have a low-mass companion 
on an exterior orbit. Most of these companions have been scattered onto wide orbits beyond $\sim1$ au, and 
no giant has two companions as does WASP-47b. This system therefore seems to have undergone disc migration.

\section{Dynamical effects on planetary system multiplicities}

As we saw above, a giant planet intruding into a close-in system may be ejected, but its intrusion may result in 
the loss of one or more of the inner planets. Similarly, in many of our systems where the giant coexists with 
the inner system, the number of planets in the inner system has been reduced through induced instability. This 
motivates an investigation of the extent to which the dynamics of an outer planetary system can affect the 
multiplicity of \textit{Kepler}-detectable systems close to the star.

This question of multiplicity is of particular interest as many studies suggest a ``\textit{Kepler} dichotomy'' 
between systems of high multiplicity and singles \citep[e.g.,][]{Johansen+12,FangMargot12}. This may be a signature 
of formation \citep{Dawson+15,MoriartyBallard15}, internal instability \citep{PuWu15,VolkGladman15} or 
externally-induced instability \citep{Mustill+16a}. We explore the latter scenario here.

Simulations suggest that for every hot Jupiter formed via 
high-eccentricity migration, at least ten times as many fail, either colliding with the star or 
failing to acquire the necessary high eccentricity \citep[e.g.,][]{Anderson+15}. This suggests that a few 
tens of percent of outer systems may be dynamically active, and we investigate the effects of this on 
close-in multiple systems. We explore two dynamical scenarios: scattering in systems of multiple giant planets
\citep[e.g.,][]{RF96,Chatterjee+08} and Kozai perturbations by a distant binary
\citep[e.g.,][]{WuMurray03,FabryckyTremaine07}.

For each of our scenarios we run 300 simulations each with a triple \textit{Kepler} system drawn from a debiased 
population where its probability of being chosen is inversely proportional to the probability of seeing a triple 
transit. For our Kozai runs we assign each system a single outer giant planet between 1 and 10 au, and a stellar 
binary companion between 50 and 1000 au. The planets are on circular, near-coplanar orbits while the binaries 
have an isotropic inclination distribution and eccentricities drawn uniformly between 0 and 1. For our 
planet--planet scattering runs we place in each system four giant planets beyond 1 au with orbits squeezed 
sufficiently close to ensure instability within the short 10 Myr time of the integrations \citep{Chatterjee+08}. We incorporate 
post-Newtonian relativistic forces into \textsc{Mercury} as these can significantly affect the dynamics, 
particularly for the Kozai scenario.

\subsection{Results}

\begin{figure}[t]
  \centering
  \hspace{-1cm}
  \includegraphics[width=12cm]{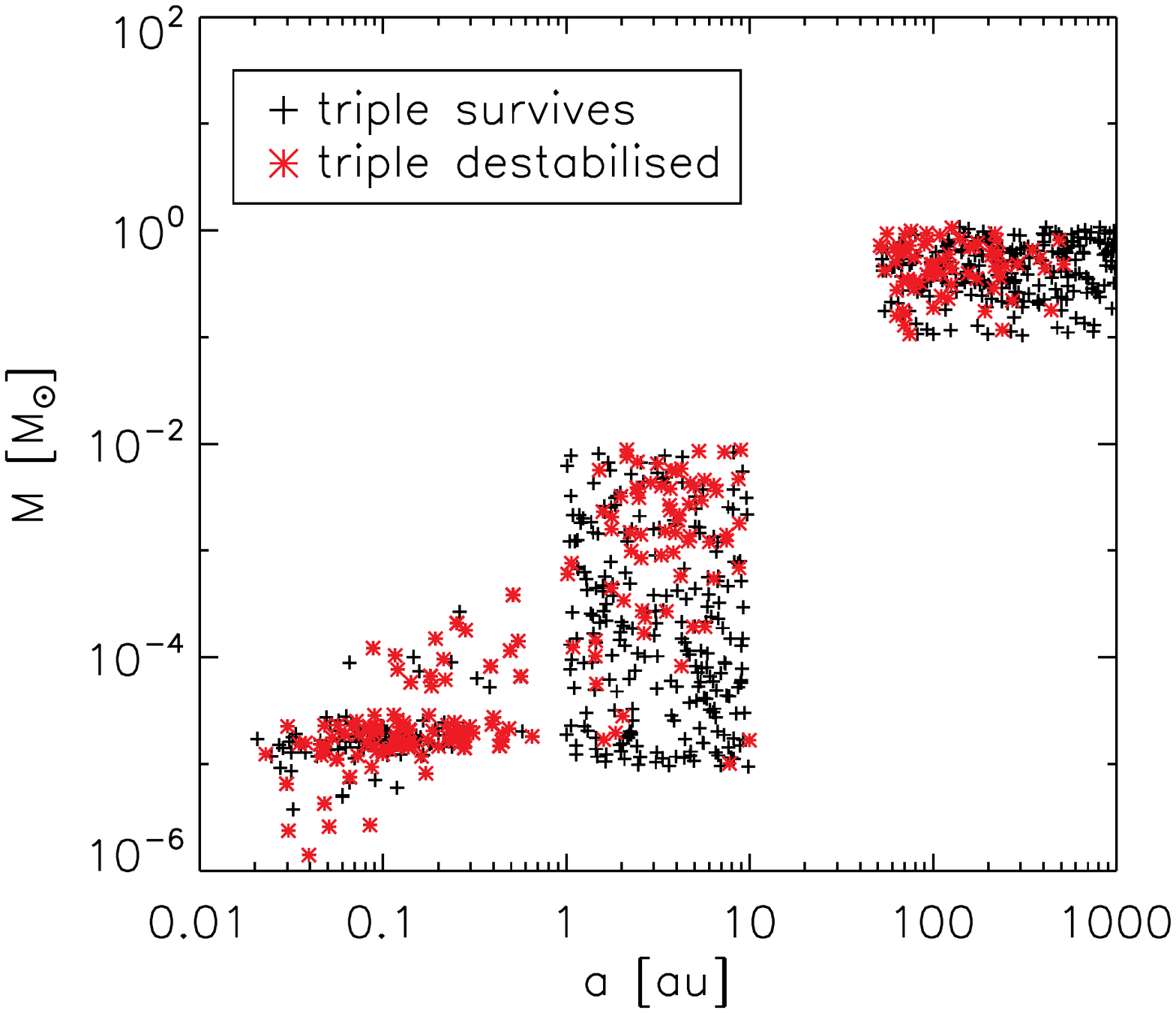}
  \caption{Effects of Kozai-perturbed giant planets on \textit{Kepler} triple systems. Each integration 
    comprises a triple system within 1 au, a giant planet between 1 and 10 au, and a binary star between 
    50 and 1000 au. Systems where one or more of the inner triple were lost are marked in red; those where
    they all survived, in black.}
  \label{fig:a-m}
\end{figure}

In each set of simulations, we destabilise around 25\% of our close-in triples systems. 
For our binary runs our initial set of 300 \textit{Kepler} triples is reduced to 230 after 10 Myr, while for our 
 scattering runs the number of survivors is 224. The survival of these systems is shown in Figure~\ref{fig:a-m} 
for the binaries, where we see that destabilisation is more likely with more massive 
outer planets and in tighter binaries. In the binary case the inner planets are destabilised at roughly the 
same rate at all semi-major axes, whereas in the scattering case the wider inner systems are more likely to be 
disrupted. This is likely because the binary perturbations can drive planets to smaller pericentres more 
easily than can planet--planet scattering.

Although the overall rate of destabilisation is similar in both scenarios, the number of surviving inner planets 
in the destabilised systems differs. 34 of the 70 destabilised inner systems in the binary runs lost all their inner 
planets, with 11 being reduced to two and 25 to one planet. On the other hand, in the scattering runs only 9 out 
of 85 inner systems lost all their planets, while 34 were reduced to two and 33 to one. This is again likely due 
to the weaker perturbations felt in the scattering systems as the outer planets' pericentres are harder to force 
to low values.

Further analysis of these cases is ongoing \citep{Mustill+16a}.  We are studying the excitation of mutual inclinations 
amongst the inner planets, as well as investigating the possible role of destructive planet--planet collisions 
\citep{Mustill+16b}.

\section{Conclusion}

We see that the multiplicities of close-in planetary systems can be significantly affected by the dynamics 
of outer systems. This is seen very strongly in hot Jupiter systems, where high-eccentricity migration 
almost always results in the complete destruction of any pre-existing planets at $\sim0.1$ au. The lack of 
close companions to most currently known hot Jupiters---with WASP-47b being the only exception---therefore 
currently supports a high-eccentricity, dynamical pathway for their migration. This will be tested further 
in the future as new space missions (CHEOPS, TESS, PLATO) will search many more hot Jupiter systems for 
close companions.

The effects of outer planets that fail to become hot Jupiters can also be significant. Our preliminary 
simulations reveal that $\sim25\%$ of dynamically active outer systems would reduce the multiplicities 
of any inner planetary systems during planet--planet scattering or Kozai cycles. Thus, the 
destabilising effects of  
outer planets on inner systems may go some way towards resolving the ``\textit{Kepler} dichotomy'', 
although other effects are likely to play a role. The true extent of the disruptive influence of outer planets 
on inner systems will be clarified as the outer regions of planetary systems are studied by Gaia, GPI, 
SPHERE, and longer-baseline RV surveys.

\acknowledgments{This work is funded by grant number KAW 2012.0150 from the Knut and Alice Wallenberg foundation, the Swedish Research Council (grants 2010-3710 and 2011-3991), and the European Research Council starting grant 278675-PEBBLE2PLANET. The author wishes to thank the conference organisers for a stimulating and enjoyable meeting.}



\bibliography{mustill_biblio}   
\bibliographystyle{aa} 

\end{document}